\documentclass[aps,prl,twocolumn,epsfig,preprintnumbers,superscriptaddress,10pt]{revtex4}

\usepackage{graphicx}
\usepackage{dcolumn}
\usepackage{bm}

\newcommand{\be}{\begin{equation}}
\newcommand{\ee}{\end{equation}}
\newcommand{\bea}{\begin{eqnarray}}
\newcommand{\eea}{\end{eqnarray}}
\newcommand{\ba}{\begin{eqnarray}}
\newcommand{\ea}{\end{eqnarray}}

\newcommand{\beq}{\begin{equation}}
\newcommand{\eeq}{\end{equation}}
\newcommand{\beqa}{\begin{eqnarray}}
\newcommand{\eeqa}{\end{eqnarray}}
\newcommand{\beqar}{\begin{eqnarray*}}
\newcommand{\eeqar}{\end{eqnarray*}}

\def\pT{{p_{\rm T}}}
\def\phiR{{\phi_{\rm R}}}

\begin{document}

\preprint{CERN-PH-TH/2009-226}

\title{Eccentricity fluctuations make flow measurable  in high multiplicity p-p collisions}

\author{Jorge Casalderrey-Solana} 
\author{Urs Achim Wiedemann}

\affiliation{    Physics Department, 
    Theory Unit, CERN,
    CH-1211 Gen\`eve 23, Switzerland}


\begin{abstract}
Elliptic flow is a hallmark of collectivity in hadronic collisions. Its measurement relies
on analysis techniques which require high event multiplicity and could be applied
so far to heavy ion collisions only. Here, we delineate the conditions under which
elliptic flow becomes measurable in the samples of high-multiplicity
($dN_{\rm ch}/dy \geq 50$) p-p collisions, which will soon be collected
at the LHC. We observe that fluctuations in the p-p interaction region can result in
a sizable spatial eccentricity even for the most central p-p collisions. Under relatively
mild assumptions on the nature of such fluctuations and on the eccentricity
scaling of elliptic flow, we find that the resulting elliptic flow signal in high-multiplicity
p-p collisions at the LHC becomes measurable with standard techniques.

\end{abstract}

\maketitle


In high-energy hadronic collisions, all projectiles, be it protons or nuclei, have finite
spatial extension and thus collide generically at finite impact parameter. This finite
impact parameter, as well as event-by-event fluctuations, both result in an azimuthally 
asymmetric shape of the interaction region in the plane 
transverse to the beam direction. If particle production is not governed solely
by independent local processes, it can be modified collectively by the 
azimuthally asymmetric spatial gradients present  in the collision geometry. 
In this case, the spatial eccentricity $\epsilon$ of the interaction region results in
a momentum anisotropy of single inclusive particle momentum distributions $dN$ 
in the azimuthal angle $\phi$, which can be characterized by the 
harmonic flow coefficients $v_n$
\begin{eqnarray}
	\frac{dN}{dy\, d{\bf p_T}} &=& \frac{1}{2\pi} \frac{dN}{p_T\, dp_T\, dy}
	          \left[ 1 + 2 \sum_n v_n\, \cos n\left(\phi - \phi_R\right) \right]\, .
	          \nonumber \\
	  v_n(\pT, y) &\equiv& \left.\left<\cos n\left(\phi-\phiR \right) \right> \right|_{\pT, y}\, .
\end{eqnarray}
In general, the $v_n'$s depend on transverse momentum $\pT$ and 
rapidity $y$. The measurement of finite harmonic flow coefficients $v_n$ 
is widely regarded as one of the most direct dynamical manifestations of 
collective behavior in high-energy hadronic collisions 
\cite{Ollitrault:1992bk,Poskanzer:1998yz}. 

The measurement of flow coefficients $v_n$ is complicated by the fact that
it requires information about the azimuthal orientation $\phi_R$ of the reaction plane. 
As we recall below, whether this reaction plane is experimentally accessible 
depends on both, the strength of the flow signals $v_n$ and the multiplicity 
$n_{\rm mult}$ in the phase space window in which $v_n$ is determined. 
This is so since microscopic dynamics such as dijet production is a source of 
azimuthal asymmetry uncorrelated to $\phi_R$ and needs to be disentangled 
from the collectively preferred macroscopic motion of all particles, characterized 
by $v_n$. Up until now, a sufficiently large signal strength at sufficiently high event multiplicity
has been observed only in heavy ion collisions, and measurements of flow 
coefficients $v_n$ have not been reported for p-p collisions. It is the main
purpose of this Letter to provide generic arguments for why at the LHC collective
flow coefficients $v_n$ may become experimentally accessible for the first time in 
proton-proton collisions
and to discuss how this measurement would impact the understanding of soft physics both 
in p-p and in heavy ion collisions.

There are several, mutually compatible standard techniques to discriminate
collective flow coefficients from non-flow 
effects~\cite{Ollitrault:1993ba,Poskanzer:1998yz, Borghini:2001vi}. 
Here, we focus on the cumulant analysis \cite{Borghini:2001vi}  which is based
on the observation that azimuthal correlations between arbitrary particle pairs $(i,j)$
in an event,
\be
\left<e^{in\left(\phi_i-\phi_j\right)}\right>=v_n v_n + \delta\, ,
\ee
do not arise solely from global correlations with $\phi_R$ (which give rise to
$v_n$), but that they can also have other, microscopic origins $\delta$. Since the 
non-flow effects parametrized by $\delta$ are, by definition, correlated at most to a small
subset of all particles in the event, they are parametrically suppressed by one 
power of $n_{\rm mult}$. Hence, the 
elliptic flow coefficient $v_2$ can 
be reliably extracted from two-particle correlations if 
$v_2\lbrace 2 \rbrace > 1/\sqrt{n_{\rm mult}}$. The assumption $v_n^2 \gg \delta$ 
that measured azimuthal two-particle correlations are dominated by a global correlation 
with $\phi_R$ can then be tested systematically by studying higher-order particle
correlations. For instance, one can analyze four-particle correlations (4th order cumulants)
which require only a signal size $v_2\lbrace 4 \rbrace > 1/\left(n_{\rm mult}\right)^{3/4}$. 
By going to higher cumulants, one achieves at best a sensitivity 
$v_2 > 1/n_{\rm mult}$. In general, the cumulant analysis allows one to
  disentangle the collective flows $v_n$ from non-flow corrections by 
systematically exploiting the different multiplicity scaling of flow and non-flow
contributions in different $n_p$-particle correlations.

With such standard techniques, the lowest order collective flow coefficients, in particular
$v_1$, $v_2$ and $v_4$, have been characterized in heavy ion collisions unambiguously 
and over a 
wide range of impact parameter at center of mass energies $\sqrt{s_{\rm NN}}$ between 
2 and 200 GeV~\cite{Alt:2003ab,Adams:2003am, Adler:2003kt,Back:2004mh,Ollitrault:1997vz}. 
 In the following, we focus mainly on elliptic flow 
$v_2$ at mid-rapidity. Odd harmonic coefficients vanish in this case by symmetry and $v_2$
is known to parametrize the dominant momentum space asymmetry. The size of $v_2$
measured in experiments at the Relativistic Heavy Ion Collider 
RHIC are very large. For instance, for transverse momenta in the range $p_T \simeq 2$ 
GeV, semi-peripheral Au-Au collisions at RHIC result in more than twice as many hadrons produced in the direction of the reaction plane, than orthogonal to it 
($v_2(p_T \sim 2\, {\rm GeV}) > 0.2$). The transverse momentum integrated $v_2$
reaches $v_2 \simeq O(0.1)$. Remarkably, the experimental praxis of $v_2$-measurements
indicates that the parametric bounds on
$v_2\lbrace 2 \rbrace,\, v_2\lbrace 4 \rbrace, \, ...\, v_2\lbrace n \rbrace$   
given above provide realistic
 numerical estimates for the feasibility of $v_2$ measurements
if one uses for $n_{\rm mult}$ values of order of the charge multiplicity per unit
rapidity $dN_{\rm ch}/dy$. We summarize this information
 in Table~\ref{bounds}  for $n_{\rm mult} =$ 30, 50 and 80. 
These values of $n_{\rm mult}$ are smaller than $dN_{\rm ch}/dy$ 
in sufficiently central heavy (Au) collisions, but they are comparable to 
the values in semi-peripheral collisions of lighter (Cu) nuclei. 

\begin{table}[h]
\begin{ruledtabular} 
\begin{tabular}{ccc}
&$n_p=2$&$n_p=4$
\\
\hline
$v_2 \,\,(n_{\rm mult}=30)$&$> 0.18 $&$ >0.09$ 
\\
\hline
$v_2 \,\,(n_{\rm mult}=50)$&$> 0.14 $&$ >0.05$ 
\\
\hline
$v_2 \,\,(n_{\rm mult}=80)$&$> 0.11 $&$ >0.04$ 
\end{tabular}
\end{ruledtabular}
\caption{Estimates of the minimal signal strength $v_2\{ n_p\}$, which can be discriminated
from non-flow effects in an $n_p$-cumulant analysis based on $n_{\rm mult}$
particles.}
\label{bounds}
\end{table}

The interpretation of elliptic flow measurements in heavy ion collisions relies on
the observation that $v_2$ is correlated with the initial spatial eccentricity $\epsilon$ 
of the transverse overlap region of the two projectiles~\cite{Ollitrault:1992bk} ,
\be
\label{ste}
\epsilon
=\frac{\left<y'^2\right>-\left<x'^2\right>}{\left<y'^2\right>+\left<x'^2\right>} \,.
\ee
Here, averages are performed with respect to the matter distribution right
after the collision, and $x'$ and $y'$ denote the lengths along the main axis 
of an ellipsoid describing this distribution. 
For sufficiently central heavy (Au) and lighter (Cu) ion collisions, it is  found that 
 $v_2 \propto \epsilon$~\cite{Alver:2006wh,Alver:2008zza}. 
Remarkably, this is a generic expectation of fluid dynamic simulations of heavy ion 
collisions. Models of ideal dissipation-free hydrodynamics, which by construction describe  
collision scenarios of maximal collective flow, can account quantitatively for the
size of the elliptic flow measured at 
RHIC~\cite{Teaney:2001av,Huovinen:2001cy,Hirano:2002ds} and
the expected dissipative corrections are anomalously 
small~\cite{Teaney:2003kp,Luzum:2008cw,Dusling:2007gi,Song:2007fn}.
In conjuction with this interpretation, the observation of very large $v_2$-signals 
\cite{Adams:2003am,Adler:2003kt,Back:2004mh} is arguably one of the most far 
reaching discoveries of the RHIC heavy ion program.

\begin{figure}
\centering
\includegraphics[width=1.0\linewidth]{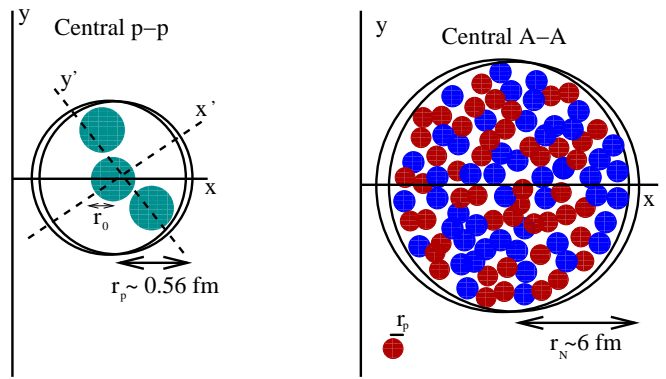}
\vskip -0.3 cm
\caption{\label{spots} Schematic view of region of hadron production may 
be located in the transverse overlap region of a central proton-proton and central
nucleus-nucleus collision respectively. Depending on the number and size of
hadronically active regions, large eccentricities can result even in central collisions.
}
\vskip -0.5 cm
\end{figure}

We now turn to the question whether elliptic flow may be measurable in p-p
collisions at the LHC. 
In p-p collisions studied so far, one may explain the apparent absence of an elliptic flow 
signal by pointing to the fact that  the $dN_{\rm ch}/dy$ in these
collisions is too low to make $v_2$ measurable (see Table~\ref{bounds}).
However, while 
Monte Carlo simulations for minimum bias $dN_{\rm ch}/dy$ distributions in 
$\sqrt{s} = 14$ TeV p-p collisions peak at low values $< 10$ for the 
non-diffractive contribution, they show a pronounced high-multiplicity tail,
typically reaching values as high as $dN_{\rm ch}/dy \sim 60$. Despite their
model dependence, these simulations strongly indicate that abundant samples of 
high-multiplicity p-p events with $dN_{\rm ch}/dy \geq 50$ will be measured at the LHC.
Such a multiplicity is comparable to that reached in semi-peripheral 
(centrality class 40 - 60 \%) Cu-Cu collisions at $\sqrt{s_{\rm NN}} = 62.4$ GeV at RHIC,
and for these latter collisions elliptic flow 
has been measured. Whether elliptic flow is also measurable in high-multiplicity 
p-p event sample at the LHC then depends on the signal strength $v_2$ and on the
relative strength of non-flow corrections in p-p collisions. 

 To estimate the strength of the elliptic flow signal $v_2$, we now discuss the
 initial spatial eccentricity $\epsilon$ of hadronic collisions. In the collisions of  
heavy (Au or Pb) ions, this eccentricity is determined solely by the transverse spatial 
overlap. More precisely, in a nucleus-nucleus collisions $dN_{\rm ch}/dy$ scales 
approximately with the average number $N_{\rm part}$ of participant nucleons,
which scales with the area of the nuclear overlap. Therefore, selecting 
a multiplicity class in A-A amounts to selecting on impact parameter and 
determines the shape of the nuclear overlap region.
For large $N_{\rm part}$, it is reasonable to make the smoothness assumption that 
the interactions between 
the $N_{\rm part}$ nucleons result in a homogeneous density distribution within the 
area of the nuclear overlap (for illustration, see right hand side of Fig.~\ref{spots}). 
If this assumption would carry over to p-p collisions, then the highest multiplicity p-p 
 collisions would be the most central ones, their spatial eccentricity would be close to zero,
 and so would be the flow signal $v_2 \propto \epsilon$.  Previous estimates of the
 magnitude of $v_2$ were based on this smoothness 
 assumption~\cite{d'Enterria:2009hd,Luzum:2009sb} or on other 
 methods~\cite{Cunqueiro:2008uu} and reported small, non-measurable values. 
  
  However, sizable deviations from the smoothness assumption have been found 
  in modeling lighter (Cu) ion collisions~\cite{Alver:2006wh,Alver:2008zza}. In these 
  systems, the relatively small number of 
  nucleon-nucleon interactions results in event-by-event fluctuations
  of the density distribution which can increase the initial spatial eccentricity significantly 
  above its geometric estimate. The predicted eccentricity scaling of $v_2$ 
  is only confirmed in Cu-Cu collisions, if these fluctuations are taken into 
  account~\cite{Alver:2006wh,Alver:2008zza}. 
  
  For p-p collisions, the $N_{\rm part}$-scaling of event multiplicity used in A-A
  does not apply. It remains a priori unclear to what extent a cut on event multiplicity
  amounts to a cut on impact parameter. MC simulations of the underlying event in p-p
  do assume indeed a correlation between impact parameter and the number of independent
  partonic interactions $N_{\rm MPI}$, which determines event multiplicity and, depending
  on model assumptions, may be as large as $N_{\rm MPI}^{\rm max} \leq 30 - 80$
  \cite{Sjostrand:2004pf}.
  However, the spatial distribution of hadronic activity does not enter the
  dynamics of these simulations. It is conceivable that all hadronic activity, even if
  emerging from a large number $N_{\rm MPI}$ of partonic interactions, is located in
  few 'hot spots' in the transverse plane. Such a picture has been advocated 
  previously in several contexts. For instance, models which view the proton as a collection of 
  three black disks of diameter $d$ result in a total p-p cross section,
 which matches experimental data for a surprisingly small diameter of 
 $d \simeq 0.2\, {\rm fm}$. Such simple constituent quark  models 
  account well for gross features of hadronic collisions, such as the ratios of hadronic cross
  sections (e.g. $\sigma_{\pi\, p}/ \sigma_{pp} \simeq 2/3$) over a wide range of $\sqrt{s}$. 
  A more field theoretic motivation of a picture of hot spots may be based on the fact
  that a large value of $N_{\rm MPI}$ requires high parton densities in the proton. In 
  QCD evolution equations such densities can arise from multiple branching of a few 
  partonic components of high momentum fraction - and to the extent to which these
  branchings are collinear, partons within the proton wave function will be located
  indeed in few 'hot spots'.  
  
\begin{figure}
\includegraphics[angle=0,width=0.8\linewidth]{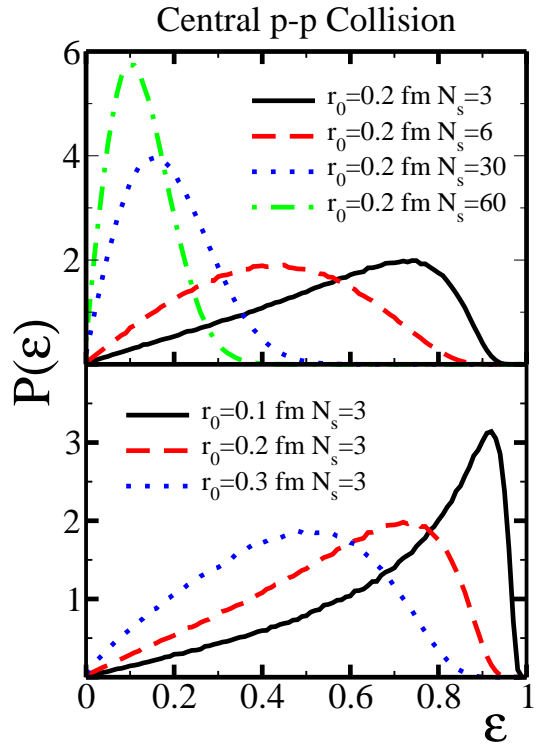}
\vskip -0.3 cm
\caption{
\label{fig:pe}
Eccentricity distribution of central high multiplicity p-p collisions (central) for different 
number of interaction regions $N_s$  (top) of different size $r_0$ (bottom). 
}
\vskip -0.5 cm
\end{figure}

 Irrespective of a specific dynamical picture, these considerations prompt us to consider
 scenarios for which the entire hadronic activity in a proton-proton collision is localized
 in $N_s$ interactions regions of radius $ r_0$. Within each interaction region, 
density is distributed homogenous with Gaussian profile, and the regions are 
distributed randomly according to the density profile of the proton (see left panel of 
Fig.~\ref{spots} for $N_s =3$). To calculate the eccentricity for this class of models, 
we take into account that for finite $N_s$ the relevant eccentricity
is the "participant eccentricity" 
\cite{Alver:2008zza}
\be
\label{parte}
\epsilon=\frac{\sqrt{\left(\sigma_y^2-\sigma_x^2\right)^2+4\sigma^2_{xy}}}{\sigma^2_y+\sigma^2_x}
\, .
\ee
Here $\sigma^2_{x}=\left\{x^2\right\} -\left\{x\right\}^2$, 
$\sigma^2_{y}=\left\{y^2\right\} -\left\{y\right\}^2$, 
$\sigma_{xy}=\left\{x y\right\}-\left\{x\right\}\left\{y\right\}$ and the event-by-event average
$\{...\}$ is taken over
the distribution of interaction. This coincides with the definitions Eq. (\ref{ste}) 
in the limit of a homogeneous density. The resulting probability distribution for different
sizes and different numbers of interaction regions is shown in Fig.~\ref{fig:pe}.
While the eccentricity vanishes indeed in the limit of a homogeneous distribution 
($N_s \to \infty$), we find that it is sizable for a wide parameter range. 

Elliptic flow develops on an event-by-event basis.   However, the experimental determination
of $v_2$ demands averaging over events and thus over distributions of geometric
shapes.  This averaging procedure is subtle; the relevant moment of 
the distribution $P(\epsilon)$ is different for different methods of extracting 
$v_2$ \cite{Bhalerao:2006tp}. We will focus on extracting $v_2$ from 
4-particle cumulants, $v_2\{4\}$, for which $\epsilon$ scales 
with~\cite{Bhalerao:2006tp}
\be
\label{ep4}
\epsilon\{4\}\equiv \left(2 \left<\epsilon^2\right>^2 -\left<\epsilon^4\right> \right)^{1/4}\, .
\ee
Here, the average  is performed over the $P(\epsilon)$ shown in Fig.~\ref{fig:pe}. 
For $N_s=3$ our model yields
the values of $\epsilon\{4\}=0.69,\,0.54 ,\,0.40 $ for the
different values of the spot size in a zero impact parameter collision. 

The precise transfer of spatial anisotropy $\epsilon\{4\}$ to  momentum anisotropy $v_2\{4\}$ 
requires a full dynamical calculation, but its magnitude can be estimated on 
general grounds~\cite{Bhalerao:2005mm}. For dilute systems, there is experimental
evidence \cite{Alt:2003ab} that the ratio $v_2/\epsilon$
depends linearly on the transverse matter density, $v_2/\epsilon \propto 
\left(dN_{\rm ch}/dy\right) / \left<S\right>$~\cite{Drescher:2007cd}. Here, 
$\left<S\right>=4 \pi \sqrt{\sigma_x^2 \sigma_y^2-\sigma^2_{xy}}$ denotes the 
mean transverse area. For denser systems, the hydrodynamic limit is reached 
and the ratio becomes independent of the system size.
In \cite{Bhalerao:2005mm} the following interpolation formula was suggested 
\be
\label{knudsen}
v_2\{4\}= \epsilon\{4\}
\left(\frac{v_2}{\epsilon}\right)^{hydro}\frac{1}{1+\frac{\bar \lambda}{K_0} \frac{\left<S\right>} {\frac{dN}{dy}}  }
\ee
where $\left(v_2/\epsilon\right)^{hydro}$ denotes the hydrodynamic limit, and 
$K_0$ and $\bar \lambda$ are constants which depend on the microscopic dynamics. 
The combination $S \bar \lambda/dN/dy $ can be understood 
as the Knudsen number of the system \cite{Bhalerao:2005mm}.

\begin{figure}
\begin{minipage}[t]{0.5\textwidth}
\centering
\includegraphics[angle=-90,width=0.8\linewidth]{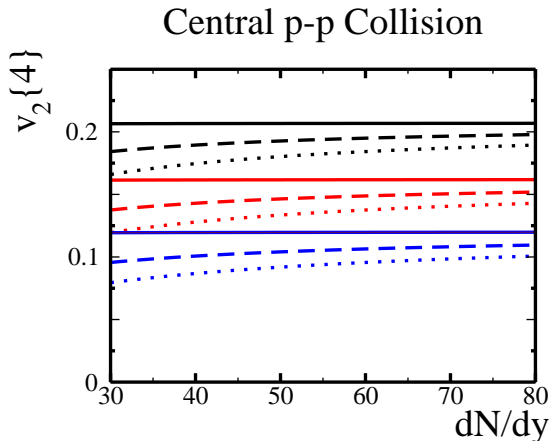}
\end{minipage}
\vskip-0.3cm
\caption{
\label{fig:v2k}
The flow signal $v_2\{4\}$ as a function of multiplicity in most 
central p-p collisions, for models of $N_s = 3$ interaction regions of  
radius $r_0=0.1$ (top 3 curves), $r_0= 0.2$ (middle curves), and 
$r_0 = 0.3$ fm (bottom curves). Signals calculated for
$\bar \lambda/K_0=0,\,5.8$ and $11.6\, {\rm fm}^{-2}$ are displayed
by  solid, dashed and dotted lines, respectively. }
\vskip -0.5 cm
\end{figure}

To estimate $v_2$ for p-p at the LHC, we assume that one can use Eq. (\ref{knudsen})
with the input $(v_2/\epsilon)^{hydro}=0.3$ and $\bar \lambda /K_0=5.8 \,{\rm fm}^{-2}$
extracted from the analysis of heavy ion collisions at RHIC~\cite{Drescher:2007cd}.
This may be motivated by observing that many collective features depend mainly
on event multiplicity and that 
the multiplicity of the p-p collisions considered here is comparable to that
in peripheral lighter (Cu) ion collisions at RHIC, for which (\ref{knudsen}) applies.
Under this assumption, we find for $N_s = 3$ 'hot spots'
the large elliptic flow values $v_2\{4\}$ displayed in Fig.~ \ref{fig:v2k}.
Also for $N_s = 6$ sufficiently small 'hot spots', the resulting $v_2\{4\}$ is 
found to be measurable (data not shown). 
To illustrate effects of theoretical uncertainties on deviations from equilibrium, we have varied 
in Fig. ~\ref{fig:v2k} the value of $\bar \lambda /K_0$ between its hydrodynamic limit 
$\bar \lambda /K_0\rightarrow 0$ and twice the fitted value quoted above.

Not measuring $v_2$ in p-p collisions can have several reasons, such as a small
spatial eccentricity of the interaction region, or viscous effects larger than expected
from Eq.~(\ref{knudsen}). However, combining the results shown in  Fig.~ \ref{fig:v2k} and 
Table \ref{bounds}, we conclude that strong but physically conceivable density 
fluctuations in the proton can lead to large values of the elliptic flow 
parameter $v_2$ in high multiplicity p-p collisions. 
Measuring elliptic flow in these collisions would provide a novel constraint on the number
and distribution of multiple parton interactions within a p-p collision,
 which is a key input in the modeling of 
the underlying  event. At the same time, measuring flow in the smallest experimentally 
accessible system would show  that collectivity develops on a sub-fermi 
time scale; since any fluid picture demands local equilibration, this measurement would provide a tight constraint  on the equilibration and dissipation mechanisms of QCD.  


\noindent 
{\bf Note added.} While preparing this letter, Ref.~\cite{Bozek:2009dw} appeared.
By studying the eccentricity for a model of two flux tubes in a p-p collision, this
work arrives at similar conclusions about the measurability of $v_2$. 

\noindent 
{\bf Acknowledgments.}
We would like to acknowledge P. Skands and G. Milhano for useful discussion.
JCS has been supported by a Marie Curie Intra-European Fellowship
 (PIEF-GA-2008-220207).


\begin{thebibliography}{99}

\bibitem{Ollitrault:1992bk}
  J.~Y.~Ollitrault,
  Phys.\ Rev.\  D {\bf 46}, 229 (1992).

\bibitem{Poskanzer:1998yz}
  A.~M.~Poskanzer and S.~A.~Voloshin,
  Phys.\ Rev.\  C {\bf 58}, 1671 (1998).

\bibitem{Ollitrault:1993ba}
  J.~Y.~Ollitrault,
  Phys.\ Rev.\  D {\bf 48}, 1132 (1993).

\bibitem{Borghini:2001vi}
  N.~Borghini, P.~M.~Dinh and J.~Y.~Ollitrault,
  Phys.\ Rev.\  C {\bf 64}, 054901 (2001).
  
\bibitem{Adams:2003am}
  J.~Adams {\it et al.}  [STAR Collaboration],
  Phys.\ Rev.\ Lett.\  {\bf 92}, 052302 (2004).

\bibitem{Adler:2003kt}
  S.~S.~Adler {\it et al.}  [PHENIX Collaboration],
  Phys.\ Rev.\ Lett.\  {\bf 91}, 182301 (2003).

\bibitem{Back:2004mh}
  B.~B.~Back {\it et al.}  [PHOBOS Collaboration],
  Phys.\ Rev.\  C {\bf 72}, 051901 (2005).

\bibitem{Alt:2003ab}
  C.~Alt {\it et al.}  [NA49 Collaboration],
  Phys.\ Rev.\  C {\bf 68}, 034903 (2003).
.  [arXiv:nucl-ex/0303001].

\bibitem{Ollitrault:1997vz}
  J.~Y.~Ollitrault,
  Nucl.\ Phys.\  A {\bf 638} (1998) 195.

\bibitem{Alver:2006wh}
  B.~Alver {\it et al.}  [PHOBOS Collaboration],
  Phys.\ Rev.\ Lett.\  {\bf 98}, 242302 (2007).

\bibitem{Alver:2008zza}
  B.~Alver {\it et al.},
  Phys.\ Rev.\  C {\bf 77}, 014906 (2008).

\bibitem{Teaney:2001av}
  D.~Teaney, J.~Lauret and E.~V.~Shuryak,
  arXiv:nucl-th/0110037.

\bibitem{Huovinen:2001cy}
  P.~Huovinen, P.~F.~Kolb, U.~W.~Heinz, P.~V.~Ruuskanen and S.~A.~Voloshin,
  Phys.\ Lett.\  B {\bf 503}, 58 (2001).

\bibitem{Hirano:2002ds}
  T.~Hirano and K.~Tsuda,
  Phys.\ Rev.\  C {\bf 66}, 054905 (2002).

\bibitem{Teaney:2003kp}
  D.~Teaney,
  Phys.\ Rev.\  C {\bf 68}, 034913 (2003).
  
\bibitem{Luzum:2008cw}
  M.~Luzum and P.~Romatschke,
  Phys.\ Rev.\  C {\bf 78}, 034915 (2008)
  [Erratum-ibid.\  C {\bf 79}, 039903 (2009)].

\bibitem{Song:2007fn}
  H.~Song and U.~W.~Heinz,
  Phys.\ Lett.\  B {\bf 658}, 279 (2008).

\bibitem{Dusling:2007gi}
  K.~Dusling and D.~Teaney,
  Phys.\ Rev.\  C {\bf 77}, 034905 (2008).

\bibitem{d'Enterria:2009hd}
  D.~d'Enterria, G.~K.~Eyyubova, V.~L.~Korotkikh, I.~P.~Lokhtin, S.~V.~Petrushanko, L.~I.~Sarycheva and A.~M.~Snigirev,
  arXiv:0910.3029 [hep-ph].

\bibitem{Luzum:2009sb}
  M.~Luzum and P.~Romatschke,
  arXiv:0901.4588 [nucl-th].
  
\bibitem{Cunqueiro:2008uu}
  L.~Cunqueiro, J.~Dias de Deus and C.~Pajares,
  arXiv:0806.0523 [hep-ph].

\bibitem{Sjostrand:2004pf}
  T.~Sjostrand and P.~Z.~Skands,
  JHEP {\bf 0403}, 053 (2004).

\bibitem{Bhalerao:2006tp}
  R.~S.~Bhalerao and J.~Y.~Ollitrault,
  Phys.\ Lett.\  B {\bf 641}, 260 (2006).

\bibitem{Bhalerao:2005mm}
  R.~S.~Bhalerao, J.~P.~Blaizot, N.~Borghini and J.~Y.~Ollitrault,
  Phys.\ Lett.\  B {\bf 627}, 49 (2005).

\bibitem{Drescher:2007cd}
  H.~J.~Drescher, A.~Dumitru, C.~Gombeaud and J.~Y.~Ollitrault,
  Phys.\ Rev.\  C {\bf 76}, 024905 (2007).

\bibitem{Bozek:2009dw}
  P.~Bozek,
  arXiv:0911.2397 [nucl-th].

\end{thebibliography}
\end{document}